\documentclass{elsart}
\usepackage{graphicx,amssymb}

\def\del{\partial}
\def\g{\gamma}
\def\ha{\frac{1}{2}}
\def\psibar{\overline{\psi}}

\begin{document}

\begin{flushright}
SLAC-PUB-9482 \\
UMN-D-02-3 \\
SMUHEP/02-01
\end{flushright}

\begin{frontmatter}

\title{The mass renormalization of nonperturbative light-front
Hamiltonian theory: An illustration using truncated, 
Pauli--Villars-regulated Yukawa interactions\thanksref{thanks1}}
\thanks[thanks1]{Work supported in part by the Department of Energy
under contract numbers DE-AC03-76SF00515, DE-FG02-98ER41087,
and DE-FG03-95ER40908.}

\author{Stanley J. Brodsky}
\address{Stanford Linear Accelerator, Stanford University,
Stanford, California 94309}

\author{John R. Hiller}
\address{Department of Physics, University of Minnesota-Duluth,
Duluth, Minnesota 55812}

\author{Gary McCartor}
\address{Department of Physics, Southern Methodist University,
Dallas, TX 75275}

\begin{abstract}
We obtain analytic, nonperturbative, approximate solutions of
Yukawa theory in the one-fermion sector using light-front quantization.
The theory is regulated in the ultraviolet by the introduction of heavy
Pauli--Villars scalar and fermion fields, each with negative norm.
In order to obtain a directly soluble problem, fermion-pair creation
and annihilation are neglected, and the number of bosonic
constituents is limited to one of either type.
We discuss some of the features of the wave function of the eigensolution,
including its endpoint behavior and spin and orbital angular momentum
content.  The limit of infinite Pauli--Villars mass receives
special scrutiny.
\end{abstract}

%

\end{frontmatter}

\section{Introduction}
\label{sec:Introduction}

It is remarkable that the first analysis of renormalization in
quantum field theory was performed in the context of a
nonperturbative problem -- the QED calculation of the Lamb Shift in
hydrogenic atoms by Bethe in 1948~\cite{Bethe:id}.
In effect, one calculates
the energy shift to order $\alpha$ and all orders in the atomic
binding from the difference~\cite{Brodsky:1966vn}:
\begin{equation} \label{eq:DeltaEn}
\Delta E_n = <\overline{\psi_n}|\Sigma(Z\alpha)-\delta m|\psi_n>\,,
\end{equation}
where $\Sigma(Z\alpha)$ is the order-$\alpha$ self energy of the
electron evaluated in the background of the nuclear Coulomb potential
$-Z\alpha\over r$, and $\delta m = \Sigma(Z = 0)$ is the free-electron
mass shift.  The $-\delta m= m_0-m$ counterterm arises from the 
order-$\alpha$ contribution to the physical mass.
As shown by Yennie and Erickson~\cite{Brodsky:1966vn}, the atomic energy
shift $\Delta E_n$ can be evaluated in terms of the expectation values of
gauge invariant operators which are functionals of the field strength 
$F_{\mu\nu}$ of the background field. The dominant term in the shift of the
$nS$ levels is of order
\begin{equation}
\Delta E_n \sim {\alpha\over \pi}(Z \alpha)^4\ln{(Z\alpha)^2} m\,.
\end{equation}
The logarithmic dependence indicates the inherently nonperturbative nature
of the renormalization problem.
The atomic-physics analysis of renormalization is appropriate to cases
such as an infinitely heavy nucleus where the binding interaction is
effectively static.  In principle, one can take into account dynamical
effects from retarded interactions and finite mass sources by using
effective interaction methods~\cite{Blokland:2001fn}.

Relativistic systems in quantum field theory can be analyzed
by quantization on the light front~\cite{Dirac} in a manner resembling
nonrelativistic theory.  The eigenstates of the invariant light-front
Hamiltonian $H_{\rm LF} = P^-P^+-P^2_\perp$ satisfy the LF Heisenberg equation
\begin{equation}
H_{\rm LF} |\psi_n> = M^2_n |\psi_n>\,.
\end{equation}
The eigenvalue problem can be rewritten in matrix form by introducing a
free Fock basis. The matrix equation becomes discrete in momentum
space by introducing periodic boundary conditions, as in
the discrete light-cone quantization (DLCQ) 
method~\cite{PauliBrodsky,DLCQreview}.
The theory can be rendered ultraviolet finite by
introducing Pauli--Villars (PV) ghost
fields~\cite{pv,bhm1,bhm2,bhm3,bhm4,Paston:1997hs,Paston:2000fq}.  
We have recently demonstrated the viability of
this type of PV regularization by applying it to (3+1)-dimensional 
Yukawa theory of spin-half fermions and scalar bosons and have 
obtained nonperturbative DLCQ solutions
of this theory in low particle number sectors~\cite{bhm3}.
We have also been able to solve such theories
analytically in the limit of exact degeneracy of the  negative 
and positive norm states~\cite{bhm4}.
Infrared divergences do not appear in neutral bound states such as
color-singlet hadrons in QCD, since the gauge-particle interactions 
cancel at long wavelength.

The interaction terms in $H_{\rm LF}$ change particle number and contain the
quantum effects associated with dressing the constituents analogous to
$\Sigma(Z\alpha)$. The counterterms arising from the difference of the
physical and bare mass of the massive constituents give a subtraction
term~\cite{Weisberger:hk} analogous to the subtraction term in 
Eq.~(\ref{eq:DeltaEn}), which is
\begin{equation}
\delta M^2_n = 
    - \sum_q <\overline{\psi_n}|{\delta m_q^2\over x_q}|\psi_n>\,,
\end{equation}
where ${1\over x_q} = {P^+\over k^+_q}$ is evaluated within the sum over
Fock states.

In practice, carrying out the above analysis is complicated in
nonperturbative relativistic problems by the necessity that the Fock
state has to be truncated.  However, we do not perform sector-dependent
renormalization~\cite{SectorDependent}, 
and instead retain the bare parameters of the original
Lagrangian, the fermion mass and the coupling, as the only parameters
to be adjusted.  The shift in the fermion mass is the only counterterm.
Contributions to the mass shift from instantaneous fermion interactions
do not appear if one includes a negative-metric fermion as part of the
Pauli--Villars regularization~\cite{bhm4}, 
since these instantaneous terms are mass independent.

In this paper we shall carry out the above nonperturbative mass
renormalization program in the context of the LF Hamiltonian formulation
of Yukawa theory. The interactions of a massive fermion and light scalar
boson generate an effective fermion bound state, for which we are able
to obtain approximate, but analytic, nonperturbative solutions.  A set of
negative-norm heavy scalar and fermion Pauli--Villars fields is used to
regulate the ultraviolet divergences while preserving the chiral
symmetry of the perturbative mass shift. In order to obtain a directly soluble
problem, fermion-pair creation and annihilation are neglected, and the
number of bosonic constituents is limited to one of either type.
We shall discuss some of the features of the
wave function of the eigensolution, including its endpoint behavior and
spin and orbital angular momentum content.  The dependence
of the renormalized theory on the mass of the Pauli--Villars fields
has some unexpected features which we discuss in detail.

In principle, the light-front wave functions of QCD can be computed directly
by the diagonalization of the light-front Hamiltonian.   The DLCQ
method~\cite{PauliBrodsky,DLCQreview} provides a discretization scheme
which transforms the eigenvalue problem of QCD into the problem
of diagonalizing very large sparse matrices.  Although the Fock space is
truncated,  the DLCQ method  retains the essential Lorentz symmetries of the
theory including boost independence.  DLCQ has also provided an important tool
for analyzing string and higher dimension theories~\cite{stringtheories}.
The DLCQ method has been successfully applied to QCD and other gauge theories
in one space and one time dimensions~\cite{DLCQreview}.   There have also been
applications of DLCQ to (3+1)-dimensional non-gauge theories~\cite{DLCQreview}.

Application to QCD$_{3+1}$ is computationally
intensive~\cite{Hollenberg}, because of the large numbers of degrees of 
freedom; however, the solution of the bound-state eigenvalue problem 
corresponding to the hadronic spectrum would be a very important step.
Given the projection of the
eigensolutions on the light-front Fock basis, one can compute
observables such as form factors and transition
amplitudes~\cite{Brodsky:1980zm,Brodsky:1998hn}, the underlying
features of deep inelastic scattering structure functions, the
distribution amplitudes which control leading twist contributions
to hard exclusive processes~\cite{Brodsky:1989pv}, and the skewed
parton distributions which can be measured in deeply virtual
Compton scattering~\cite{Brodsky:2000xy,Diehl:2000xz}.
First-principle computations of exclusive decay amplitudes of
heavy hadrons, such as the $D$ and $B$
mesons~\cite{Keum:2000wi,Beneke:1999br}, require knowledge of
heavy and light hadron wave functions in order to extract the
phases and other parameters of the electroweak theory.
Light-front techniques can also be applied to traditional
nuclear physics~\cite{Miller}. The light-front representation is
boost-independent and provides the nonperturbative input and matrix
elements required for such analyses.

An important feature of
light-front Hamiltonian is the simplicity of spin and angular momentum
projections: the sum rule for the angular momentum of the eigensolution
$J^z =\sum^n_{i=1} S^z_i + \sum^{n-1}_{i=1} L^z_i $
holds Fock state by Fock state.  Here the sum is
over the spin projections $S_z$ of the constituents in the $n$-particle Fock
state.  There are only $n-1$ contributions to the internal orbital angular
momentum.  The spin projections also provide a convenient way to classify
independent contributions to the wave functions.
Our truncation to two partons limits $L^z_i$ to the values 0 and $\pm 1$.

In Sec.~\ref{sec:Framework} we discuss the Yukawa Hamiltonian and its
regularization and renormalization, as well as the Hamiltonian eigenvalue
problem, which we solve, given the truncation to two particles.
Since the methods we use are analytic, we are able to find two-parton
solutions in the continuum theory without DLCQ or other discretization.
We discuss the nature of the solutions in two limits, one in 
Sec.~\ref{sec:Equal}
where the PV masses are equal and another in Sec.~\ref{sec:Unequal} where the
PV boson mass approaches infinity more slowly than the PV fermion mass.
We are particularly interested in the chiral properties, the large
transverse momentum fall-off, and the end-point behavior of the eigensolutions
in the light-cone variables $x_i= k^+/P^+$ of the constituents.
In Sec.~\ref{sec:NoLimit} we argue that in calculations where the
representation space is truncated, it is necessary to keep the values of 
the PV masses finite even in cases where it is computationally possible
to take the limit of infinite PV masses.  It is possible that some of the 
effects we see may be related to the triviality of Yukawa 
theory~\cite{triviality}, but similar considerations probably 
apply to asymptotically free theories.
Section~\ref{sec:Discussion} contains our conclusions.

Our calculations are
somewhat similar to those of Bylev, G{\l}azek, and Przeszowski~\cite{Glazek},
except that they did not use a covariant regulation
procedure; it is the effect of the covariant regulator that will form the
focus of our discussion here.  Similar work in a purely scalar theory
has been done by Bernard {\em et al}.~\cite{Bernard}.
For a more formal treatment of dressed
constituents, see~\cite{EffectiveFermions}.

The notation that we use for light-cone coordinates is
\begin{equation}
x^\pm = x^0+x^3\,,\;\;
\vec{x}_\perp=(x^1,x^2)\,.
\end{equation}
The time coordinate is $x^+$, and the dot
product of two four-vectors is
\begin{equation}
p\cdot x=\frac{1}{2}(p^+x^- + p^-x^+)
                -\vec{p}_\perp\cdot\vec{x}_\perp\,.
\end{equation}
The momentum component conjugate to $x^-$ is $p^+$,
and the light-cone energy is $p^-$.
Light-cone three-vectors are identified by
underscores, such as
\begin{equation}
\underline{p}=(p^+,\vec{p}_\perp)\,.
\end{equation}
For additional details, see Appendix A of Ref.~\cite{bhm1}
or the review~\cite{DLCQreview}.

\section{Computational Framework}
\label{sec:Framework}

Taking the physical fermionic and bosonic fields to be $\psi_1$
and $\phi_1$, respectively, and the PV
(negative-metric) fields to be $\psi_2$ and $\phi_2$, the
Yukawa action becomes
\begin{eqnarray}
S=\int d^4x&&
\left[\ha(\del_\mu\phi_1)^2-\ha\mu_1^2\phi_1^2
-\ha(\del_\mu\phi_2)^2+\ha\mu_2^2\phi_2^2\right.
\nonumber \\
+&& \frac{i}{2}\left(\psibar_1\g^\mu\del_\mu-(\del_\mu\psibar_1)\g^\mu\right)
     \psi_1-m_1\psibar_1\psi_1 \nonumber \\
&& \left.
-\frac{i}{2}\left(\psibar_2\g^\mu\del_\mu-(\del_\mu\psibar_2)\g^\mu\right)
      \psi_2+m_2\psibar_2\psi_2
-g\phi\psibar\psi\right]\,,
\end{eqnarray}
where the scalar three-point interaction is expressed in terms of zero-norm fields
\begin{equation}
\psi \equiv (\psi_1 + \psi_2)\,,  \quad
                \phi \equiv (\phi_1 + \phi_2)\,.
\end{equation}
For simplicity, we will not consider a $\phi^4$ term; with pair
creation removed, it will  not be required.

If the bare mass of the fermion is zero,
the bare action (the same as above but with the PV fields set equal to zero)
possesses a discrete chiral symmetry -- invariance under the
transformations $\psi \rightarrow \gamma_5 \psi$, $\phi \rightarrow -\phi$.
One consequence of this symmetry is that the physical mass is also zero.  The
full action, including the PV fields breaks this symmetry explicitly.  There
are two possible versions of this breaking:  if we require that the PV Fermi
field is not transformed, the cross term in the interaction breaks the
symmetry; if we require the PV Fermi field to transform in the same way as the
physical field, the PV mass term breaks the chiral symmetry.
Thus an interesting question, which we will examine below, is whether chiral
symmetry will be restored in the limit of large PV masses where the
unphysical states decouple.   It will be a point of interest to see if the
symmetry is restored, at least in the sense that the bare mass and the
physical mass are proportional to each other.

The corresponding light-cone Hamiltonian, except for the addition
of the PV fields, has been given by McCartor and Robertson~\cite{mr}.
Here we include the PV fields but neglect
pair terms and any other terms which involve anti-fermions.
The resulting Hamiltonian is
\begin{eqnarray} \label{eq:HLC}
P^-&=&
   \sum_{i,s}\int d\underline{p}
      \frac{m_i^2+{\vec p}_\perp^2}{p^+}(-1)^{i+1}
          b_{i,s}^\dagger(\underline{p}) b_{i,s}(\underline{p}) \\
   && +\sum_{j}\int d\underline{q}
          \frac{\mu_j^2+{\vec q}_\perp^2}{q^+}(-1)^{j+1}
              a_j^\dagger(\underline{q}) a_j(\underline{q}) \nonumber \\
   && +\sum_{i,j,k,s}\int d\underline{p} d\underline{q}\left\{
      \left[ V_{-2s}^*(\underline{p},\underline{q})
             +V_{2s}(\underline{p}+\underline{q},\underline{q})\right]
                 b_{j,s}^\dagger(\underline{p})
                  a_k^\dagger(\underline{q})
                   b_{i,-s}(\underline{p}+\underline{q})\right.  \nonumber \\
      &&\left.  +\left[U_j(\underline{p},\underline{q})
                    +U_i(\underline{p}+\underline{q},\underline{q})\right]
               b_{j,s}^\dagger(\underline{p})
                a_k^\dagger(\underline{q})b_{i,s}(\underline{p}+\underline{q})
                    + h.c.\right\}\,,  \nonumber
\end{eqnarray}
where
\begin{equation}
U_j(\underline{p},\underline{q})
   \equiv \frac{gm_j}{\sqrt{16\pi^3}}\frac{1}{p^+\sqrt{q^+}}\,,\;\;
V_{2s}(\underline{p},\underline{q})
   \equiv \frac{g}{\sqrt{8\pi^3}}
      \frac{\vec{\epsilon}_{2s}^{\,*}\cdot\vec{p}_\perp}{p^+\sqrt{q^+}}\,,
\end{equation}
$m_1$ is the mass of the bare fermion,
$\mu\equiv\mu_1$ is the physical boson mass,
$\mu_2$ and $m_2$ are the masses of the PV boson and fermion,
respectively, and
${\vec \epsilon}_{2s}\equiv-\frac{1}{\sqrt{2}}(2s,i)$.  
The $V$ interaction
introduces one unit of relative orbital angular momentum projection $L_z$
which is compensated by the change in fermion spin projection $S_z$ to 
conserve $J_z.$~\cite{Brodsky:2000ii}.  The nonzero commutators are
\begin{eqnarray}
\left[a_i(\underline{q}),a_j^\dagger(\underline{q}')\right]
          &=&(-1)^i\delta_{ij}
            \delta(\underline{q}-\underline{q}')\,, \nonumber \\
   \left\{b_{i,s}(\underline{p}),b_{j,s'}^\dagger(\underline{p}')\right\}
     &=&(-1)^i\delta_{ij}   \delta_{s,s'}
            \delta(\underline{p}-\underline{p}')\,.
\end{eqnarray}
The Fock-state expansion for a spin-1/2 fermion eigenstate
of the Hamiltonian is
\begin{eqnarray}
\Phi_\sigma&=&\sum_{n_1,n_2,k_1,k_2=0}^\infty
   \prod_{n=1}^{n_{\rm tot}}
  \int d\underline{p}_n  \sum_{s_n}
   \prod_{k=1}^{k_{\rm tot}}
     \int d\underline{q}_k
       \delta(\underline{P}-\sum_{n}^{n_{\rm tot}}\underline{p}_n
                     -\sum_{k}^{k_{\rm tot}}\underline{q}_k)       \\
   &  & \times
     \phi_{\sigma s_n}^{(n_i,k_j)}(\underline{p}_n;\underline{q}_k)
     \frac{1}{\sqrt{\prod_i n_i!\prod_j k_j!}}
        \prod_n^{n_{\rm tot}}  b_{i_n,s_n}^\dagger(\underline{p}_n)
      \prod_k^{k_{\rm tot}} a_{j_k}^\dagger (\underline{q}_k)|0\rangle \,,
\nonumber
\end{eqnarray}
where $n_1$ is the number of bare fermions, $n_2$ the number of
PV fermions, $k_1$ the number of physical bosons, and $k_2$
the number of PV bosons.  The total number of fermions is
$n_{\rm tot}=n_1+n_2$, and the total number of
bosons is $k_{\rm tot}=k_1+k_2$.  The $n$-th constituent fermion
is of type $i_n$, and the $k$-th boson is of type $j_k$.
This Fock state
expansion will be used to solve the eigenvalue problem
$(P^+P^-{\vec P}^2_\perp)\Phi_\sigma=M^2\Phi_\sigma$.  The normalization of
the eigenstate is
\begin{equation} \label{eq:normalization}
\Phi_\sigma^{\prime\dagger}\cdot\Phi_\sigma
=\delta(\underline{P}'-\underline{P})\,.
\end{equation}

Our first approximation will be to truncate the expansion to two particles:
\begin{equation}
\Phi^{(2)}_+=\sum_i z_i b_{i+}^\dagger(\underline{P})|0\rangle
   +\sum_{ijs}\int d\underline{l} f_{ijs}(\underline{l})
             b_{i,s}^\dagger(\underline{P}-\underline{l})
              a_j^\dagger(\underline{l})|0\rangle
\end{equation}
and reduce the eigenvalue problem by projecting onto Fock sectors.
Without loss of generality, we consider only the $J_z=+1/2$ case.
The resulting coupled equations determine the wave functions
$f_{ij\pm}$ to be
\begin{eqnarray}
f_{ij+}(\underline{l})&=&
   \frac{P^+}{M^2-\frac{m_i^2+l_\perp^2}{1-l^+/P^+}
                -\frac{\mu_j^2+l_\perp^2}{l^+/P^+}}
\nonumber \\
&&\times
\left[(\sum_k (-1)^{k+1}z_k)U_i(\underline{P}-\underline{l},\underline{l})
      +\sum_k (-1)^{k+1}z_kU_k(\underline{P},\underline{l})\right]\,,
\nonumber \\
f_{ij-}(\underline{l})&=&
   \frac{P^+}{M^2-\frac{m_i^2+l_\perp^2}{1-l^+/P^+}
                -\frac{\mu_j^2+l_\perp^2}{l^+/P^+}}
(\sum_k (-1)^{k+1}z_k)V_+^*(\underline{P}-\underline{l},\underline{l})\,,
\end{eqnarray}
corresponding respectively to Fock components where the fermion constituent is
aligned or anti-aligned with the total spin $J_z.$
The nonperturbative physics is thus contained in the determination of 
$z_i$ and $M^2$.  When the two-body wave functions
$f$ are eliminated from the one-fermion projections, we obtain\footnote{Note
that the projection onto the opposite  spin is automatically zero because the
integrand is linear in
$\vec{l}_\perp$.}
\begin{eqnarray} \label{eq:onefermion}
(M^2-m_i^2)z_i &=&
 g^2\mu_1^2 (z_1-z_2)J+g^2 m_i(z_1m_1-z_2m_2) I_0
\nonumber \\
  &&+g^2\mu_1[(z_1-z_2)m_i+z_1m_1-z_2m_2] I_1\,,
\end{eqnarray}
with
\begin{eqnarray}
I_n&=&\int\frac{dy dl_\perp^2}{16\pi^2}
   \sum_{jk}\frac{(-1)^{j+k}}{M^2-\frac{m_j^2+l_\perp^2}{1-y}
                                   -\frac{\mu_k^2+l_\perp^2}{y}}
   \frac{(m_j/\mu_1)^n}{y(1-y)^n}\,, \\
J&=&\int\frac{dy dl_\perp^2}{16\pi^2}
   \sum_{jk}\frac{(-1)^{j+k}}{M^2-\frac{m_j^2+l_\perp^2}{1-y}
                                   -\frac{\mu_k^2+l_\perp^2}{y}}
   \frac{(m_j^2+l_\perp^2)/\mu_1^2}{y(1-y)^2}\,.
\end{eqnarray}
These integrals are not independent; a change of variable
to $w=\frac{m_j^2+l_\perp^2}{1-y}+\frac{\mu_k^2+l_\perp^2}{y}$
and an interchange of integration order can be used to show
that
\begin{equation} \label{eq:JproptoI0}
J=\frac{M^2}{\mu_1^2}I_0\,.
\end{equation}
We solve the $i=2$ case of
Eq.~(\ref{eq:onefermion}) for $\zeta\equiv z_2/z_1$, to obtain
\begin{equation} \label{eq:zeta}
\zeta=\frac{g^2\mu_1^2J+g^2\mu_1(m_1+m_2) I_1 +g^2 m_1m_2 I_0}
{M^2-m_2^2+g^2\mu_1^2 J+2g^2\mu_1m_2 I_1
  +g^2 m_2^2 I_0}\,.
\end{equation}
From the remaining $i=1$ case we solve for $g^2$.  This yields
\begin{equation} \label{eq:gofm}
g^2=-\frac{(M\mp m_1)(M\mp m_2)}{(m_2-m_1)(\mu_1 I_1\pm MI_0)}\,.
\end{equation}
There are two possible solutions for $g^2$ since the remaining
equation is quadratic in $g^2$.
Substitution into (\ref{eq:zeta}) and use of (\ref{eq:JproptoI0})
reduces $\zeta$ to the remarkably simple form
\begin{equation} \label{eq:zeta2}
\zeta=\frac{M \mp m_1}{M \mp m_2}\,,
\end{equation}
independent of $I_0$ and $I_1$.

We pause here to remark that the wave function
we have obtained in the nonperturbative calculation is very similar to the
one which we would obtain using first-order perturbation theory to perturb
about the state of one  bare, physical fermion.  The only differences are
that in perturbation theory $z_2 = 0$ and $M = m_1$.  From (\ref{eq:zeta2})
we see that as $m_2 \rightarrow \infty$,  $\zeta=z_2/z_1$ will be small
as long as
$m_1 << m_2$.  Since this last requirement is necessary if we are to
expect to restore at least approximate unitarity in the limit of large PV
masses, we will insist on it.  The only significant difference between our
nonperturbative calculation and first-order perturbation theory is that in
perturbation theory $M = m_1$ while in the nonperturbative calculation $M$ is
determined by (\ref{eq:gofm}).\footnote{In practice we will fix
$M$ as a renormalization condition and use (\ref{eq:gofm}) to restrict the
behavior of $g$ and $m_1$ as functions of $m_2$.}
The form of the wave function in terms
of the parameters is exactly the same in the perturbative and nonperturbative
calculations; only the parameters are different.

In the presence of the negatively normed constituents, we
define the ``physical wave functions'' as the coefficients of
Fock states containing only positive-norm particles.  
This can be done without ambiguity by requiring that all Fock states
be expressed in terms of the positive-norm creation operators
$b_{1s}^\dagger$ and $a_1^\dagger$ and the zero-norm
combinations $b_s^\dagger\equiv b_{1s}^\dagger+b_{2s}^\dagger$
and $a^\dagger\equiv a_1^\dagger+a_2^\dagger$.  Because
$b_s^\dagger$ is null, a fermion created by
$b_s^\dagger$ is annihilated by the generalized electromagnetic
current $(\psibar_1+\psibar_2)\g^\mu(\psi_1+\psi_2)$ appropriate
to this PV-regulated theory; thus the null fermions do not
contribute to current matrix elements and should not make
a physical contribution to a state.  By analogy, $a^\dagger$
is also deemed to create unphysical contributions.
The procedure, then, is to express the wave function in terms the operators
$b_{1s}^\dagger$, $b_s^\dagger$, $a_1^\dagger$ and $a^\dagger$ acting on the
vacuum.  Any term containing a $b_s^\dagger$ or an $a^\dagger$ is then
discarded when constructing the physical state.  
This procedure is a non-orthogonal projection onto the physical
subspace.  

After application of this procedure to our case,
the physical state with spin $J_z=+1/2$ in the two-particle truncation is
\begin{eqnarray}
\Phi_{+ {\rm phys}}^{(2)}&=&(z_1-z_2)b_{1+}^\dagger(\underline{P})|0\rangle
\nonumber \\
&&+\sum_{i,j,s}\int d\underline{q} (-1)^{i+j}f_{ijs}(\underline{q})
     b_{1s}^\dagger(\underline{P}-\underline{q})a_1^\dagger(\underline{q})
              |0\rangle\,.
\end{eqnarray}
The normalization condition (\ref{eq:normalization}) fixes $z_1$.
In addition to fixing the physical mass, one additional
renormalization condition is needed.  In previous papers~\cite{bhm1,bhm2,bhm3}
we have specified a value for the expectation value in the state
\begin{equation}
\langle :\!\!\phi^2(0)\!\!:\rangle
   \equiv\Phi_\sigma^\dagger\!:\!\!\phi^2(0)\!\!:\!\Phi_\sigma .
\end{equation}
For some of the solutions given below this quantity diverges
even after renormalization, so this is not a suitable condition.
In the rest of the paper it is not necessary to specify the final renormalization
condition.  In place of specifying the final normalization condition, we
will examine features of the solution.
We will look for cases where the structure functions are finite and nonzero.  

The normalization of $\Phi_{+ {\rm phys}}^{(2)}$ and the definition
of $\langle :\!\!\phi^2(0)\!\!:\rangle$ reduce to
\begin{eqnarray} \label{eq:normint2}
1&=&(z_1-z_2)^2+\sum_s\int d\underline{l}
     \left|\sum_{ij}(-1)^{i+j}f_{ijs}(\underline{l})\right|^2\,, \\
\langle :\!\!\phi^2(0)\!\!:\rangle&=&
\sum_s\int d\underline{l} \frac{2}{l^+/P^+}
     \left|\sum_{ij}(-1)^{i+j}f_{ijs}(\underline{l})\right|^2\,.
\end{eqnarray}
These relations can be written more explicitly in terms of the
following integrals:
\begin{eqnarray}
\tilde{I}_0&=&\int \frac{dy}{16\pi^2} i_0(y)\,, \;\; 
   \tilde{I}'_0=\int \frac{dy}{16\pi^2} \frac{2}{y} i_0(y)\,, 
\\ \nonumber
\tilde{I}_1&=&\int \frac{dy}{16\pi^2} i_1(y)\,, \;\; 
   \tilde{I}'_1=\int \frac{dy}{16\pi^2} \frac{2}{y} i_1(y)\,, 
\\  \nonumber
\tilde{J}_0&=&\int \frac{dy}{16\pi^2} j_0(y)\,, \;\; 
   \tilde{J}'_0=\int \frac{dy}{16\pi^2} \frac{2}{y} j_0(y)\,, 
\\ \nonumber
\tilde{J}_1&=&\int \frac{dy}{16\pi^2} j_1(y)\,, \;\; 
   \tilde{J}'_1=\int \frac{dy}{16\pi^2} \frac{2}{y} j_1(y)\,, 
\end{eqnarray}
where
\begin{eqnarray} \label{eq:ijIntegrals}
i_0(y)&=&\int dl_\perp^2
   \left(\sum_{jk}\frac{(-1)^{j+k}}{M^2-\frac{m_j^2+l_\perp^2}{1-y}
                                   -\frac{\mu_k^2+l_\perp^2}{y}}\right)^2
   \frac{\mu_1^2}{y}\,, 
\\ \nonumber
i_1(y)&=&\int dl_\perp^2
   \left(\sum_{jk}\frac{(-1)^{j+k}m_j}
                        {M^2-\frac{m_j^2+l_\perp^2}{1-y}
                                   -\frac{\mu_k^2+l_\perp^2}{y}}\right)
\\ \nonumber && \times
   \left(\sum_{jk}\frac{(-1)^{j+k}}{M^2-\frac{m_j^2+l_\perp^2}{1-y}
                                   -\frac{\mu_k^2+l_\perp^2}{y}}\right)
   \frac{\mu_1}{y(1-y)}\,, 
\\ \nonumber
j_0(y)&=&\int dl_\perp^2
   \left(\sum_{jk}\frac{(-1)^{j+k}m_j}{M^2-\frac{m_j^2+l_\perp^2}{1-y}
                                   -\frac{\mu_k^2+l_\perp^2}{y}}\right)^2
   \frac{1}{y(1-y)^2}\,, 
\\ \nonumber
j_1(y)&=&\int dl_\perp^2
   \left(\sum_{jk}\frac{(-1)^{j+k}}{M^2-\frac{m_j^2+l_\perp^2}{1-y}
                                   -\frac{\mu_k^2+l_\perp^2}{y}}\right)^2
   \frac{l_\perp^2}{y(1-y)^2}\,.
\end{eqnarray}
For the normalization and for
$\langle :\!\!\phi^2(0)\!\!:\rangle$, we then obtain
\begin{eqnarray}
\frac{1}{z_1^2}&=&(1-\zeta)^2[1+g^2(\tilde{J}_0+\tilde{J}_1)] \nonumber \\
&& +g^2\frac{(m_1-\zeta m_2)^2}{\mu_1^2}\tilde{I}_0
      +2g^2(1-\zeta)\frac{m_1-\zeta m_2}{\mu_1}\tilde{I}_1\,, \\
\langle :\!\!\phi^2(0)\!\!:\rangle&=&
    g^2 z_1^2\left\{(1-\zeta)^2[\tilde{J}'_0+\tilde{J}'_1]\right. \nonumber \\
&& \left.+\frac{(m_1-\zeta m_2)^2}{\mu_1^2}\tilde{I}'_0
      +2(1-\zeta)\frac{m_1-\zeta m_2}{\mu_1}\tilde{I}'_1\right\}\,.
\end{eqnarray}

The boson structure functions are given by
\begin{equation}
f_{Bs}(y)\equiv \int d^2l_\perp
   \left|\sum_{ij}(-1)^{i+j}f_{ijs}(yP^+,{\vec l}_\perp)\right|^2\,.
\end{equation}
In terms of integrals already defined in (\ref{eq:ijIntegrals})
we obtain
\begin{eqnarray}
\label{eq:sfns1}
f_{B+}(y)&=&\frac{g^2z_1^2}{16\pi^2}\left[(1-\zeta)^2j_0(y)+
\frac{(m_1-\zeta m_2)^2}{\mu_1^2} i_0(y) \right. \nonumber \\
&& \left.
    +2(1-\zeta)\frac{(m_1-\zeta m_2)}{\mu_1}i_1(y)\right]\,, \\
\label{eq:sfns2}
f_{B-}(y)&=&\frac{g^2z_1^2}{16\pi^2}(1-\zeta)^2j_1(y)\,.
\end{eqnarray}

As an alternative renormalization condition one could use
the radius $R$ of the dressed-fermion state, as defined by
the slope of the Dirac form factor $F_1$.  These quantities are
related by the standard expression $R=\sqrt{-6F'_1(0)}$.
The slope can be computed from the eigenfunction 
$\Phi_{+ {\rm phys}}^{(2)}$ as
\begin{eqnarray}
-\frac{R^2}{6}=F'_1(0)&=&z_1^2\frac{g^2}{16\pi^2}
\sum_{i'j'}(-1)^{i'+j'} \sum_{ij}(-1)^{i+j}
   \int_0^1 \alpha(\alpha-1) d\alpha y^3dy 
\\ \nonumber
&& \times \left\{ \left[
  (1-\zeta)^2m_im_{i'}+(m_i+m_{i'})(1-\zeta)(m_1-\zeta m_2)(1-y) 
   \right.  \right.
\\ \nonumber
&&  \left.  \rule{1.5in}{0in}
     +(m_1-\zeta m_2)^2(1-y)^2\right]/(D_{ij}^{i'j'} )^2
\\  \nonumber
&& \left. \rule{0.25in}{0in}
               +2(1-\zeta)^2/D_{ij}^{i'j'} \right\}\,,
\end{eqnarray}
with 
$D_{ij}^{i'j'}\equiv\alpha(ym_{i'}^2+(1-y)\mu_{j'}^2)
    +(1-\alpha)(ym_i^2+(1-y)\mu_j^2)-y(1-y)M^2$.
Similarly we can extract the axial coupling~\cite{SchlumpfBrodsky}
\begin{eqnarray}
g_A&=&z_1^2(1-\zeta)^2+z_1^2\frac{g^2}{16\pi^2}
\sum_{i'j'}(-1)^{i'+j'} \sum_{ij}(-1)^{i+j} \int_0^1 d\alpha ydy 
\\ \nonumber
&& \times \left\{ \left[
  (1-\zeta)^2m_im_{i'}+(m_i+m_{i'})(1-\zeta)(m_1-\zeta m_2)(1-y) \right. \right.
\\  \nonumber
&&  \left.  \rule{2in}{0in}
  +(m_1-\zeta m_2)^2(1-y)^2  \right]
                                   /D_{ij}^{i'j'}
\\  \nonumber
&& \left. \rule{0.25in}{0in}
             +(1-\zeta)^2\log[2D_{ij}^{i'j'}] \right\}
\end{eqnarray}
and the anomalous magnetic moment $\kappa = F_2(0)$ of the dressed fermion
\begin{eqnarray}
\kappa&=&2Mz_1^2(1-\zeta)\frac{g^2}{16\pi^2}
\sum_{i'j'}(-1)^{i'+j'} \sum_{ij}(-1)^{i+j} \int_0^1 d\alpha y^2dy 
\\ \nonumber
&& \times \left[
  (1-\zeta)\left(\alpha m_{i'}+(1-\alpha)m_i\right)
     +(m_1-\zeta m_2)(1-y) \right]/D_{ij}^{i'j'} \,.
\end{eqnarray}
The result for the anomalous moment is confirmed by
comparison with Eq.~(51) of Ref.~\cite{Brodsky:2000ii}.
If the fermion $x$ in Ref.~\cite{Brodsky:2000ii} 
is written as $1-y$ and the $M$ in the numerator
is replaced by $m_1$, as per the discussion after Eq.~(46),
the two results agree, once we drop the sum over PV particles.
Note that only the two-particle Fock state contributes since 
the anomalous moment requires a change in $L_z$ without a change in
particle number.

\begin{figure}[bhtp]
\begin{tabular}{cc}
\includegraphics[width=6.75cm]{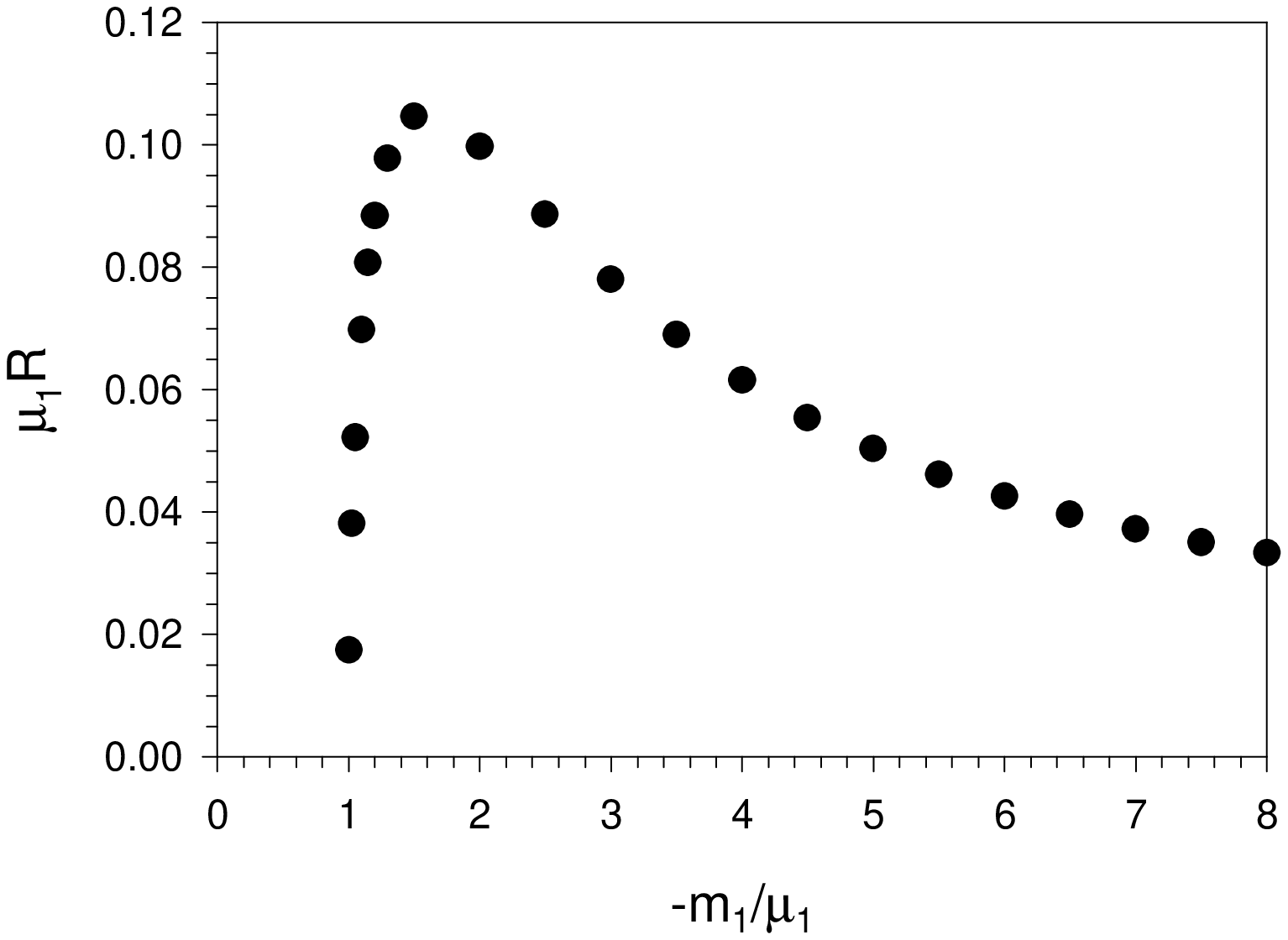} &
\includegraphics[width=6.75cm]{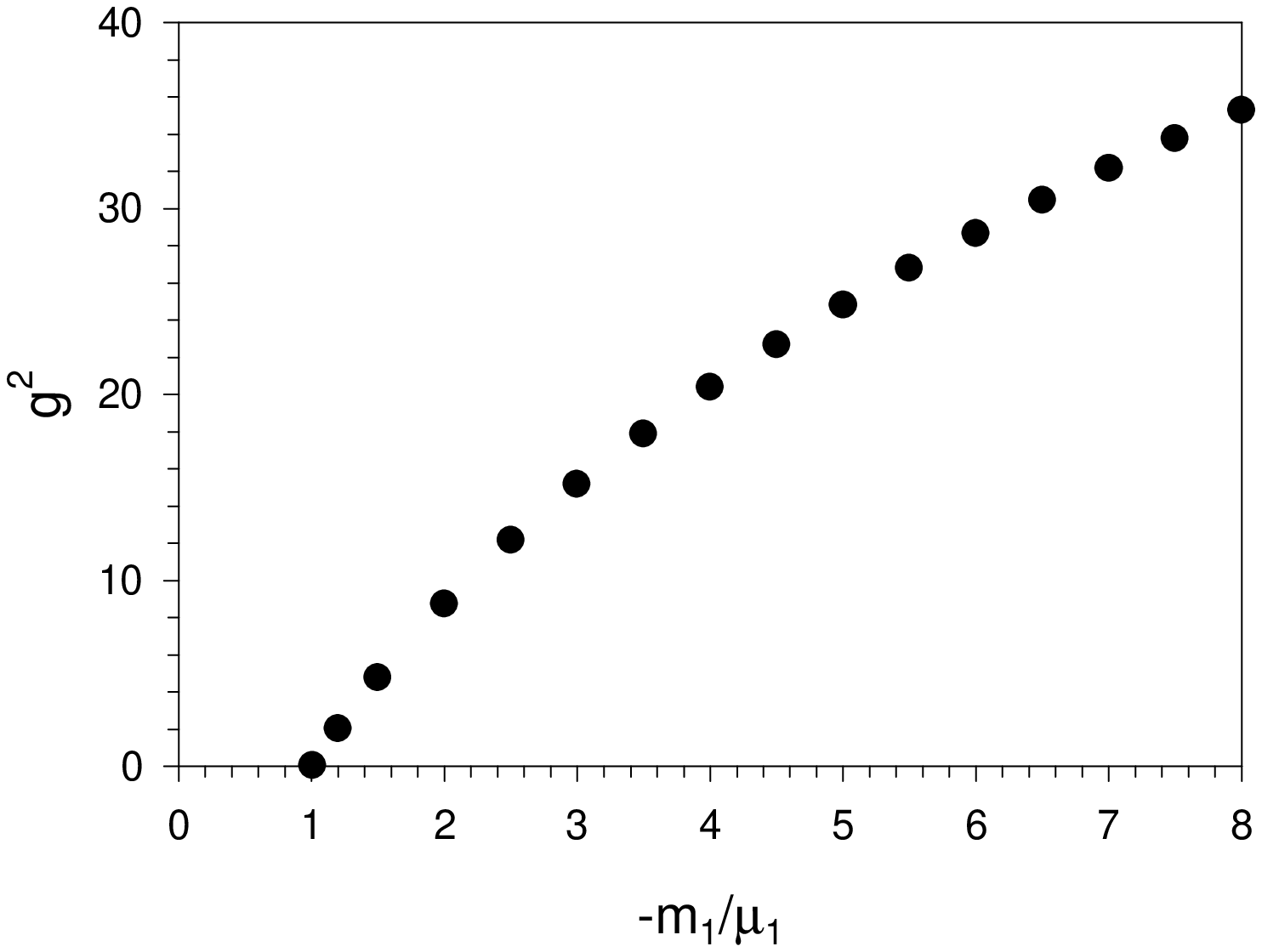} \\
(a) & (b) 
\end{tabular}
\caption{Plots of (a) the dressed-fermion radius $R$ and
(b) the bare coupling squared $g^2$ as functions of the
ratio of the bare fermion mass $m_1$ to the physical boson mass $\mu_1$.  
The PV masses are fixed at $m_2=\mu_2=10\mu_1$, and
the dressed-fermion mass at $M=\mu_1$.
\label{fig:Rg2}
}
\end{figure}

\begin{figure}[bhtp]
\centerline{\includegraphics[width=10cm]{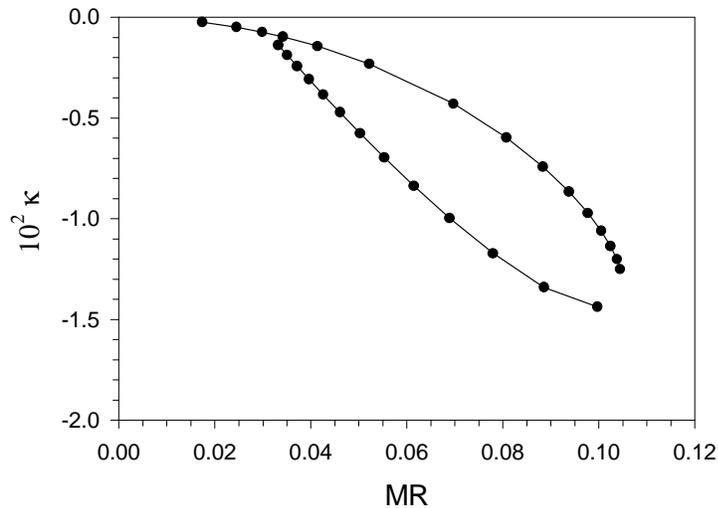}}
\caption{The anomalous moment $\kappa$ of the dressed fermion,
multiplied by $10^2$,
as a function of its radius $R$, scaled by its mass $M$.
For this particular plot the dressed-fermion mass is set equal to 
the physical boson mass $\mu_1$, and the PV masses are fixed at 
$m_2=\mu_2=10\mu_1$.  The solid lines are drawn to connect points
on the same branch.
\label{fig:kappa}
}
\end{figure}

As an example of how $R$ might be used as a renormalization
condition, we compute $R$ and $g^2$ for a series of $m_1$ values,
with $m_2=\mu_2=10\mu_1$ and $M$ fixed at $\mu_1$.\footnote{The boson 
is not dressed due to the fact that we have eliminated pair production.}
We use the lower signs in Eqs.~(\ref{eq:gofm}) and (\ref{eq:zeta2}).  The 
results are plotted in Fig.~\ref{fig:Rg2}.  The figures show that for a chosen
value of $R$ and $M$ one can obtain values for the bare 
parameters $g$ and $m_1$, with the only ambiguity being between
weak and strong coupling.
The axial coupling is essentially constant over the given range,
at a value very near unity.  The anomalous moment is plotted in
Fig.~\ref{fig:kappa}; here there is again the double-valued
structure of the radius $R$.  That the anomalous moment vanishes
as $MR\to 0$ is consistent with the Drell--Hearn--Gerasimov sum
rule, as discussed in~\cite{SchlumpfBrodsky}.

\section{Limits for large Pauli--Villars Masses}

\subsection{Equal Pauli--Villars Masses}
\label{sec:Equal}

Although the nonperturbative problem has been reduced to a
single nonlinear equation, and although all the integrals involved in that
equation can be done in closed form, the resulting expression is very long
and complex and, worse yet, is a function of many variables. 
To gain some control over the total space
in which we will look for solutions, we will fix the ratio of the
two PV masses.  A natural choice seems to be $m_2 = \mu_2$, especially if we
choose $M$, the physical fermion mass, to be equal to $\mu_1$, the physical
boson mass.

When the PV masses $m_2$ and $\mu_2$ are equal, and we make the assumption
that $m_1 << m_2$, the integrals $I_0$ and
$I_1$, multiplied by $16\pi^2$,
reduce to $\log(m_2^2/m_1^2)$ and $m_2/\mu_1$, respectively.
We therefore find from (\ref{eq:gofm}) that
\begin{equation} \label{eq:gofm2}
\frac{g^2}{16\pi^2} =-\frac{(M\mp m_1)(M\mp m_2)}
{(m_2-m_1)[m_2 \mp M \log (m_1^2/m_2^2)]}\,.
\end{equation}
With this choice of the behavior of the PV masses,
the integrals involved in the structure functions (\ref{eq:sfns1}) and
(\ref{eq:sfns2}) have no
singularities (in $m_2$) worse than logarithmic.  From (\ref{eq:gofm2}) we see
that if $m_1$ stays finite or diverges more slowly than
$m_2/\log m_2$ in the large-$m_2$ limit, 
$g$ will go to zero so fast that the structure
functions must vanish, and we will have, in that sense, a trivial theory.  A
further examination of the particular choice $m_1 \sim m_2/\log m_2$
shows that even in that case the structure functions go to zero as $m_2$ goes
to infinity.  The only choice for the behavior of $m_1$ as a function of
$m_2$ which leads to finite, nonzero structure functions is $m_1 \sim m_2$.
We therefore define $r\equiv m_1/m_2$ and hold
$r$ fixed as $\mu_2=m_2\rightarrow\infty$.  The
fractional amplitude $\zeta$ for the single-PV-fermion
state, given in (\ref{eq:zeta2}), becomes equal to $r$.
The coupling is then driven to a fixed value
\begin{equation}
\frac{g^2}{16\pi^2} =  \frac{-r}{1-r}\,.
\end{equation}
Thus $m_1$ (and $r$) must be negative.  The structure functions become
\begin{eqnarray}
\label{eq:equalfBp}
f_{B+}(y)&\simeq&  -z_1^2 r(1-r )
  \left[ 2 + y + \frac{r^2y}{1 + \left( r^2 -1\right) \,y} \right.
  \nonumber \\
&& -2\left( \frac{r \,
            \log (r^2)}{r^2-1} - \frac{y\,\log (y)}{1-y} -
         \frac{r\ \,y\,\log (r^2\,y)}
          {r^2\,y-1} \right. \nonumber \\
&& -\frac{r^2\,y\,}{1-y}\log (\frac{r^2\,y}{1 - y + r^2\,y}) +
         \frac{r\,\log (1 - y + r^2\,y)}{r^2-1} \nonumber \\
&& \left.\left.-
         \frac{ry}
            {1 +\left(r^2 -2\right) \,y} \log (\frac{1 - y + r^2\,y}{y})
                           \right) \right]\,,  \\
\label{eq:equalfBm}
f_{B-}(y)&\simeq&  z_1^2 r(1-r)( 1 - r^2)
    \left( 4y -
      \frac{\left( 1 + r^2 \right) \,
          \,y\,\log (r^2)}{r^2-1} +
      \frac{y(1+y)\log (y)}{1-y} \right. \nonumber \\
&& + \frac{y\,
         \left( 1 + r^2\,y \right) \,\log (r^2\,y)}{
         r^2\,y-1}
    + \frac{y\,\left( 1 +
         \left(2\,r^2-1 \right) \,y \right)}{1-y} \,
       \log (\frac{r^2\,y}{1 - y + r^2\,y})\nonumber \\
&&  - \frac{ \,
         \left( 2 + \left(r^2-1 \right) \,y \right) \,
         \log (1 - y + r^2\,y)}{r^2-1} \nonumber \\
&& \left. + \frac{y\,
         \left( 1 + r^2\,y \right)}
     {1+\left(r^2-2 \right) \,y} \log (\frac{1 - y + r^2\,y}{y})\right)\,. 
\end{eqnarray}
The nonorthogonal projection of the wave function ensures that 
these distributions are positive definite.  The reciprocal of
the factor $z_1^2$ is determined by the normalization condition 
(\ref{eq:normalization}) to be
\begin{eqnarray}
\frac{1}{z_1^2}&=&(1-r)^2 + 
    \frac{1}{144}\frac{g^2}{16\pi^2}
     \left[ 4( 27 - 108 r + 307 r^2) \right.  - 3\pi^2(6 - 24 r + 49 r^2) 
\nonumber \\
&& \rule{1.5in}{0in} \left. - 
                24 (3 - 12 r + 20 r^2)\ln(r^2)\right]
\end{eqnarray}
to second order in $r$.  
We thus have a one-parameter family of
theories labeled by $r$.  While the PV masses have been taken to
infinity, they have not been made infinitely large compared to the
bare fermion mass, which has been taken to minus infinity. 
The value of $g$ is finite in this limit.
We probably should not, even naively, think that all the effects of the
negatively normed states have been removed from the full solution.
To control such effects, we should consider values of $r$ which are
small in absolute value.  
Notice that $g$ is then restricted to small values.

The results of the exact solution are very different
from perturbation theory.
In first-order perturbation theory, $M$ is equal to $m_1$, 
and there is no nonlinear eigenvalue equation
and thus no restriction of the value of $g$.  Indeed, since the physical
mass, $M$, is fixed and equal to $m_1$, we could not send $m_1$ to minus
infinity as we did above.  We also note that the discrete chiral symmetry is
not restored in the large PV-mass limit.  We cannot take $m_1$ to be
zero (without obtaining a trivial theory).  We can take the physical mass,
$M$, to be zero, but that point does not occur at $m_1 = 0$.

\begin{figure}[bhtp]
\centerline{\includegraphics[width=10cm]{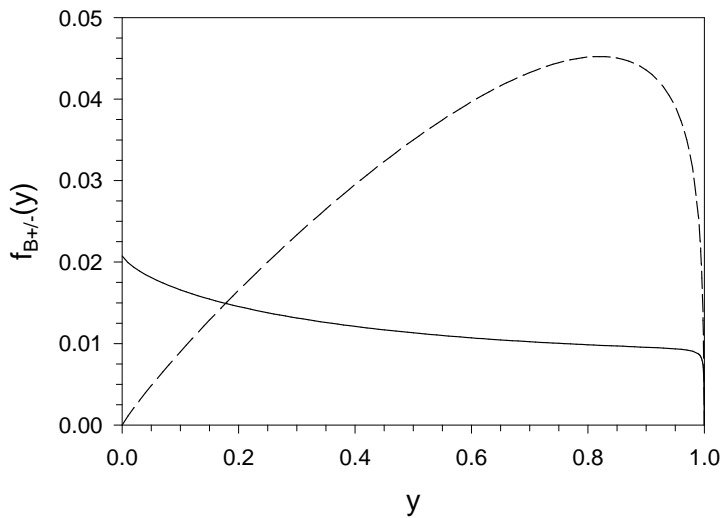}}
\caption{\label{fig:equal}
Structure functions $f_{B+}$ (solid) and $f_{B-}$ (dashed)
for the equal-PV-mass case, with $r=m_1/m_2=-0.01$,
from the forms given in Eqs.~(\ref{eq:equalfBp}) and (\ref{eq:equalfBm}) 
of the text.}
\end{figure}

Plots of the structure functions for $r=m_1/m_2=-0.01$ are given in
Fig.~\ref{fig:equal}.
We should remark on the behavior of the structure functions at the end
points.  For very large
values of the PV masses the functions are given essentially exactly by 
Eqs.~(\ref{eq:equalfBp}) and (\ref{eq:equalfBm}) for all points except 
very close to $y=0$ in the case of $f_{B+}$.  The exact structure functions
are zero at $y=0$ for all values of the PV masses; yet (\ref{eq:equalfBp})
yields a nonzero value at that point.  Thus the convergence to the
limiting forms is nonuniform. For that reason, any quantity sensitive to the
endpoint behavior, such as the expectation value of the parton light-cone
kinetic energy, should be calculated for finite values of the PV masses
and then taken to the infinite-mass limit.

\subsection{Unequal Pauli--Villars Masses}
\label{sec:Unequal}

Having obtained the results discussed in the last subsection,
one can ask whether there is any way to get results more like perturbation
theory.  As it turns out, there is: to do so we must take the limit of large
PV masses in such a way that the PV fermion mass grows much faster than the
PV boson mass.\footnote{We could let the boson mass grow as fast as $\log m_2$, 
but it is also allowed, and is simpler, to first take $m_2$ to infinity at 
finite $\mu_2$ (that limit turns out to be finite) then take $\mu_2$ to 
infinity.}

If we take the mass $m_2$ to infinity, the integrals $I_0$ and $I_1$
reduce to
\begin{equation}
16\pi^2 I_0 \simeq -\log(\mu_2/\mu_1)\,,\;\;
16\pi^2 I_1\simeq -2\frac{m_1}{\mu_1}\log(\mu_2/\mu_1)\,.
\end{equation}
The fractional amplitude $\zeta=(m_1\mp M)/m_2$ goes to zero.
When we take $m_2\rightarrow\infty$ and then take $\mu_2$ large, the eigenvalue
equation (\ref{eq:gofm}) becomes
\begin{equation} \label{eq:gofm3}
\frac{g^2}{16\pi^2} = \frac{C}{\log(\mu_2/\mu_1)}\,,
\end{equation}
where
\begin{equation} \label{eq:C}
 C =  \left(\frac{m_1 \mp M}{2m_1 \pm M}\right)\,.
\end{equation}
In this limit the structure functions reduce to
\begin{eqnarray}
f_{B+}(y)&=&\frac{g^2z_1^2}{16\pi^2}
                  \frac{(1\pm m_1/\mu_1-y)^2 y\mu_1^2}
                         {\mu_1^2 (1-y) + m_1^2 y - M^2 y (1-y)}\,, \\
f_{B-}(y)&=&\frac{g^2z_1^2}{16\pi^2}y
      \left\{  \log \left[\frac{(1 - y) \mu_2^2}
         {\mu_1^2 (1-y) + m_1^2 y - M^2 y (1-y)}\right]  -2 \right\}\,.
\end{eqnarray}

\subsubsection{$m_1$ finite}

Looking at these relations we see that if $g^2 \sim 1/\log \mu_2$,
$f_{B-}$ will be finite and nonzero while $f_{B+}$ will be zero
in the limit of large PV mass.  If $g^2$ remains finite,
$f_{B+}$ will be finite and nonzero while $f_{B-}$ will diverge,
which is untenable.  
There are two choices for the behavior of $m_1$ which will 
give us the desired behavior for $g^2$
and finite non-zero values for $f_{B-}(y)$.
One way is to choose $m_1$ to be finite and choose
its value and the signs in (\ref{eq:C}) such that
the constant $C$ is any real number we wish.
In that case, $f_{B-}$ is given by
\begin{equation}
f_{B-}(y) = 2 C z_1^2y\,.
\end{equation}
From (\ref{eq:normint2}) we find that
\begin{equation} \label{eq:z1}
  z_1^2 = \frac{1}{1 + C}\,.
\end{equation}
Thus there is a finite probability that
the state consists of a single physical fermion.  The larger the value of $C$
the smaller is that probability and the larger is the probability that the
state contains two particles.  We note, as in the case of equal PV
masses, that the discrete chiral
symmetry is not restored in the sense that if we take either $M$ or $m_1$
equal to zero, the other is not specified and disappears entirely from the
problem; the value of $C$ is fixed at either 1 or 1/2.

\subsubsection{$m_1$ proportional to $M$} \label{sec:m1proptoM}

The other possibility for the behavior of $m_1$
is to choose $m_1 \sim \pm\frac{M}{2}$ with the appropriate choice of sign
in (\ref{eq:C}).  For illustration we take the lower sign and parameterize
\begin{equation} \label{eq:m1}
m_1 = \frac{M}{2} + \frac{\mu_1}{2 c \log (\mu_2/\mu_1)}\,,
\end{equation}
with $c$ a constant.
With these choices the bare coupling constant goes to a finite value given by
\begin{equation} \label{eq:gofm4}
\frac{g^2}{16\pi^2} = \frac{3 M c}{2} \,.
\end{equation}
From this we see that $c$ should be positive.
Notice that this choice is much more like perturbation theory: instead of
$m_1 = M$, as in first-order perturbation theory,
we have (in the limit) $m_1 = \frac{M}{2}$, and the coupling
constant can be any finite number.  In this case we find for large $\mu_2$
that $z_1^2$ is given by
\begin{equation}
z_1^2 = \frac{2}{3 M c \log (\mu_2/\mu_1)} \,.
\end{equation}
There is zero probability that the system
is in the state of one physical fermion, and the entire wave function is in
the two-particle sector.  Due to the behavior of $z_1$ we find again that,
in the infinite-$\mu_2$ limit, $f_{B+}$ is zero while
\begin{eqnarray} \label{eq:sfy}
f_{B-}(y) = 2 y\,.
\end{eqnarray}

The outcome for the discrete chiral symmetry in this case is not so clear.
The fact that $m_1$ is proportional to $M$ (in the limit) suggests that it
may be restored.  On the other hand if $M$ is zero we would encounter
undefined expressions in the above derivation.  However, we can perform the 
entire calculation with $M$ set equal to zero from the start, and we find 
that if we take
\begin{equation}
m_1 =  \frac{\mu_1}{2 c \log \frac{\mu_2}{\mu_1}}\,,
\end{equation}
we obtain the structure function (\ref{eq:sfy})
and zero for $f_{B+}(y)$; this last result is in agreement with perturbation
theory.  So in that sense, the discrete chiral symmetry is restored in the
large-PV-mass limit.

We should repeat the comment of the
previous section regarding the behavior of the structure functions at the
endpoints.  For finite values of the PV masses, the structure functions
vanish at $y = 1$, but there is a nonuniform convergence.  For very large
values of the PV masses the structure function is closely proportional to $y$
for all values of $y$ except very near 1 where it falls precipitously to
zero.  In the limit of large PV masses the function converges to something
proportional to $y$ for every point except $ y=1$, where it is always
zero.  For that reason any quantity which is sensitive to the endpoint
behavior (such as the kinetic energy of the fermion or
$\langle :\!\!\phi^2(0)\!\!:\rangle$) should be calculated
for finite values of the PV masses then the limit taken.  If that exercise is
performed for $\langle :\!\!\phi^2(0)\!\!:\rangle$, we find that this quantity
does diverge.

\section{On Not Taking the Limit}
\label{sec:NoLimit}

Up to now we have taken the
limit of the PV masses going to infinity.  Here we wish to further consider
the comparison of our results with perturbation theory.  We believe that this
comparison suggests that we should not take that limit and furthermore
indicates why we should not do so.  These same considerations will suggest a
way to decide how large we should take the PV masses.

Let us fix our attention on the choices made in Sec.~\ref{sec:m1proptoM}
for taking the limit of large PV masses and fixing $m_1$, which
gave results most like perturbation theory.  The structure function
$f_{B+}(y)$ was zero in that case.  That does not happen in perturbation
theory.  Since our wave function is identical and even the parameters are
almost the same (differing only in that $m_1=M/2$), how can we get
something so different from perturbation theory?  The reason we obtained 
zero for $f_{B+}$ is
that the renormalization constant $z_1$ went to zero.  If we look at the
form of the function for finite PV mass, it is
\begin{equation}
f_{B-}(y) = \frac{g^2 [\mbox{finite quantity}]}
     {1 + g^2[\mbox{finite quantity}]
                + g^2[\mbox{finite quantity}] \log \mu_2} \,.
\end{equation}
The denominator represents $z_1^{-2}$.  Now in perturbation theory,
since the numerator is already of order $g^2$, only the 1 in the denominator
is used and the result is nonzero.  Indeed, suppose we calculate some
quantity which is finite to this order such as the anomalous magnetic
moment.  Again we would get a result of the form
\begin{equation}
\kappa = \frac{g^2 [\mbox{finite quantity}]}
   {1 + g^2[\mbox{finite quantity}] +
                  g^2[\mbox{finite quantity}] \log \mu_2} \,.
\end{equation}
If we use the methods of the previous section this quantity would again
be zero.  In perturbation theory that would not happen, again because the
divergent term in the denominator would not be used with this numerator.  Now
the divergent term in the denominator would be used in a calculation to order
$g^4$; but then there would be an order $g^4$ term in the numerator which
would cancel the divergence of the term from the denominator and give a
finite result.  That is the way perturbation theory works.  The point is
this:  we will have an accurate calculation only to the extent that the
projection of the wave function onto the excluded Fock states is small.  We
know from past calculations~\cite{bhm3} that this projection can be very
small, sometimes
even for the severe truncation we are considering here, but those results
were for finite values of the PV masses.  There will be divergences in the
excluded Fock sectors, and we must anticipate that for sufficiently large
values of the PV masses the projection of the wave function onto those
sectors will not be small.

There are two types of error associated with having finite values 
of the PV masses:
for PV masses too small we will have too much of the negative norm states in the
system.  We anticipate that such errors are approximately the larger of
$\frac{m_1}{m_P}$ or $\frac{\mu_1}{m_P}$ where $m_P$ is the smallest PV
mass.  The other type of error is a large projection of the wave function
onto the excluded Fock sectors.  That error should be approximately
\begin{equation}
 \frac{\langle \Phi_{+ {\rm phys}}^\prime|\Phi_{+ {\rm phys}}^\prime\rangle}
{\langle \Phi_{+ {\rm phys}}|\Phi_{+ {\rm phys}}\rangle}
\end{equation}
where $\Phi_{+ {\rm phys}}^\prime$
is the projection of the wave function onto the lowest excluded Fock
sector.\footnote{If some rule other than particle number is used
to truncate the space, $\Phi_{+ {\rm phys}}^\prime$ is the projection
onto the ``next'' set of vectors.}  The higher Fock wave function
$\Phi_{+ {\rm phys}}^\prime$ can be estimated
using perturbation theory, perturbing about $\Phi_{+ }$ with the projection
of $P^-$ onto the excluded sectors being chosen as the perturbing operator.
The first type of error, from negative-metric Fock states,
decreases with increasing PV mass; the second type of
error, the truncation error, will usually increase with increasing PV mass.  
Ideally we should choose the values of the PV masses to be the values where 
the two types of error are equal.  
The strategy for treating the nonperturbative system is to
include more and more of the representation space in
our calculation and to increase the value of the PV masses until
the desired accuracy is achieved.   How much of the space will be
required will depend on the problem.

\begin{figure}[bhtp]
\begin{tabular}{cc}
\includegraphics[width=6.75cm]{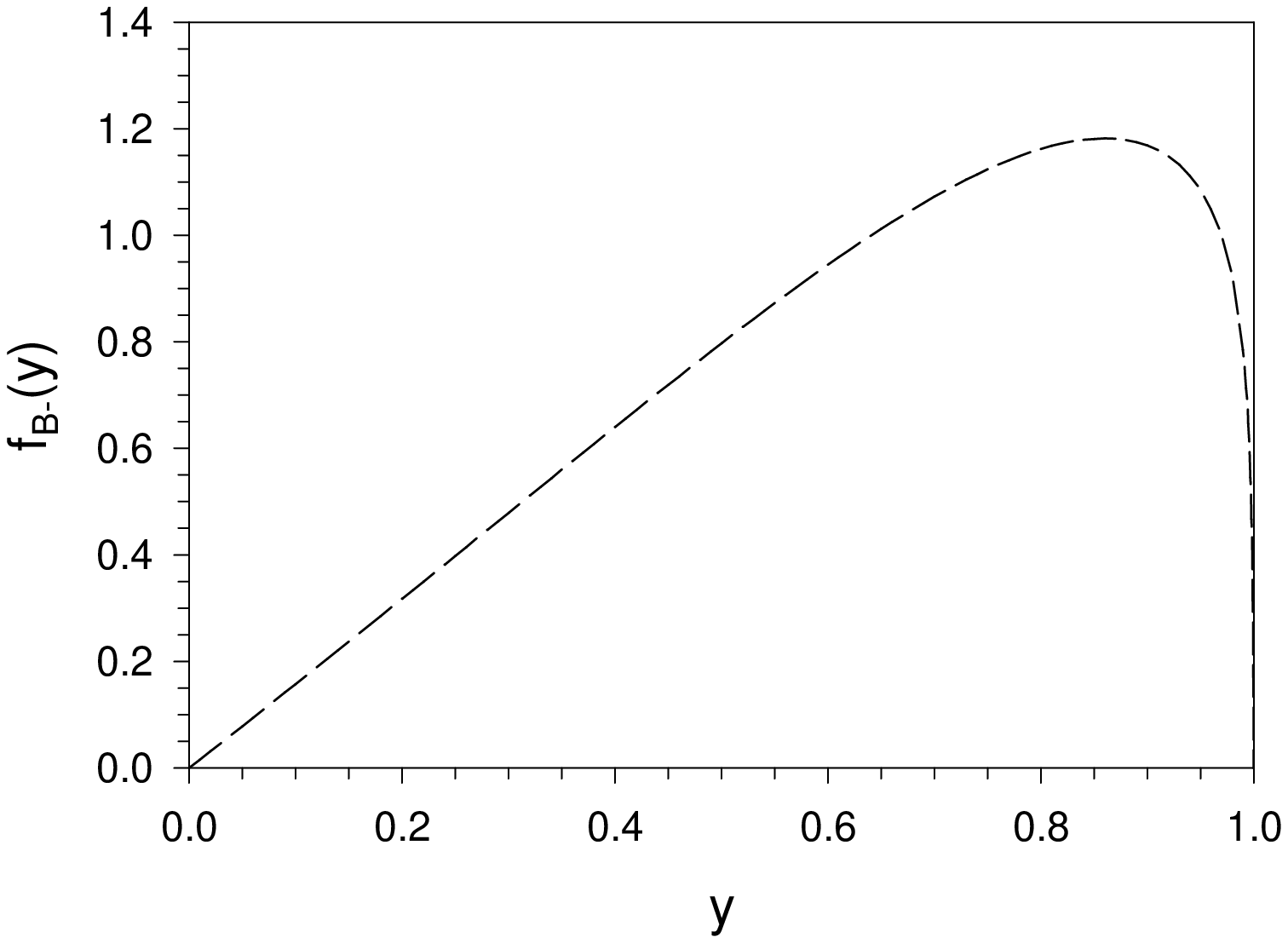} &
\includegraphics[width=6.75cm]{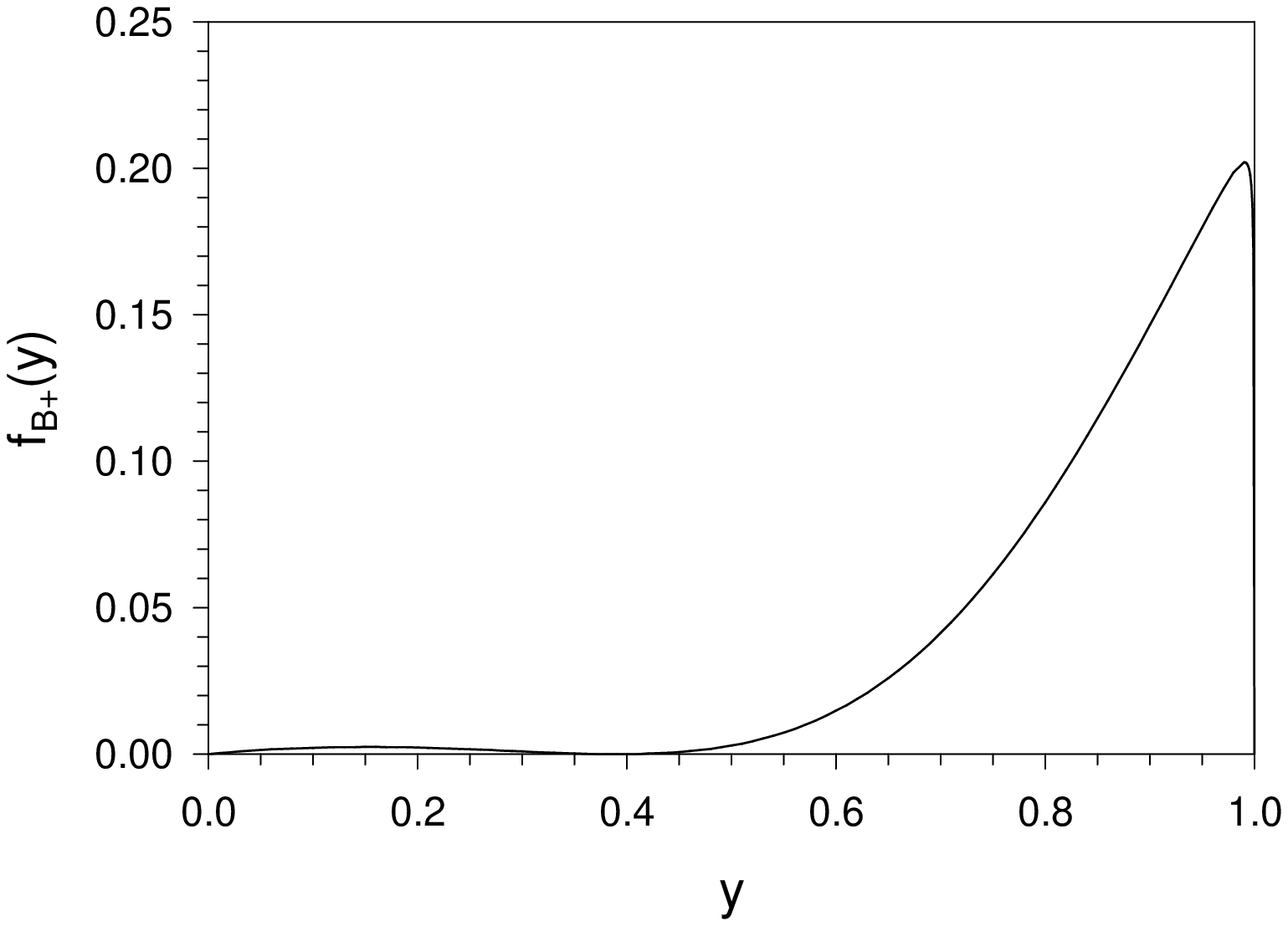} \\
(a) & (b) \\
\end{tabular}
\caption{\label{fig:unequal}
Structure functions (a) $f_{B-}$ and (b) $f_{B+}$
for the unequal-PV-mass case where $m_2\rightarrow\infty$,
$\mu_2=100\mu_1$, and $M=\mu_1$.  The bare fermion mass $m_1$
is specified by Eq.~(\ref{eq:m1}) of the text, and the coupling $g^2$
by Eq.~(\ref{eq:gofm4}), with $c=1$.  Notice that the two
plots have different vertical scales.}
\end{figure}

In later work we will attempt to make these comments
quantitative by estimating the optimum values for the PV masses.
Here we will illustrate the effects of not taking the limit, for
the trajectory in which the PV masses are taken to large values
as in Sec.~\ref{sec:m1proptoM}.  For
infinite $m_2$ but finite $\mu_2$ ($=100\mu_1$), and with the
eigenvalue equation solved by Eq.~(\ref{eq:m1}) and
Eq.~(\ref{eq:gofm4}), the structure functions, $f_{B+}$ and
$f_{B-}$ are plotted in Fig.~\ref{fig:unequal}.  These are to be
compared with the linear function (Eq.~(\ref{eq:sfy})) for
$f_{B-}$, and zero for $f_{B+}$, which result if the limit of
infinite PV masses is taken.  The structure functions for the
finite value of the mass $M$ resemble what one 
expects in a bound state.

\section{Discussion}
\label{sec:Discussion}

In this paper we have studied the regulation of Yukawa theory by the 
use of Pauli--Villars fields in such a way that the interaction is
written as a product of zero-norm fields.  Paston and Franke have 
shown that this regulation procedure gives perturbative equivalence 
with Feynman methods~\cite{Paston:1997hs}.  The theory is covariant and 
presumably finite, and there are no gauge symmetries to protect.  
Therefore, if we could solve such a theory exactly and take the limit 
of the PV masses going to infinity, the result would be the best one 
could do to give 
a meaning to the theory.  In this paper we have done our calculations 
in a severe truncation of the representation space.  Such a truncation 
violates covariance, but if the contribution of excluded Fock sectors
is sufficiently small, the consequence of the truncation is more a 
question of accuracy than of preserving symmetries: 
if we are close to the hypothetical solution mentioned above, it does 
not matter if this (small) error violates symmetries.

The reason we have used such a severe truncation is that it allows us 
to find solutions and take limits in closed form, and thus our 
interpretation of the results is not confounded by questions of 
inaccuracies introduced by numerical solutions. A significant feature of the 
calculations, which came as a surprise to us, is that the results depend strongly 
on the way in which the two PV masses are allowed to approach infinity.  
It is not true that any two different trajectories will give different results 
but rather that there are families of trajectories which give the same results. 
For instance, any trajectory on which the two PV masses are proportional to 
each other, with a fixed constant of proportionality, give the same result 
as taking the limit with the two masses set equal to each other.  Similarly, 
any path on which $\mu_2$ is logarithmically small compared with $m_2$ 
will give the same result as taking the limit $m_2 \rightarrow \infty$ 
first, then taking the limit $\mu_2 \rightarrow \infty$.  One possibility 
is that all these trajectories represent 
different phases of the theory.  Another possibility is that the effect 
is an artifact of the truncation, and, if we include more and more of the 
representation space in the calculations, the results of the various 
ways of taking the limit will approach each other.  Another possibility 
that we have considered is that some of the ways of taking the limit are 
wrong and that some principle which we have not yet discerned will determine 
the correct way to take the limit.  We hope to report further studies on 
this question in the future.

In the calculations we have given special consideration to the 
discrete chiral symmetry that is formally present in the unregulated 
Lagrangian.  Writing the interaction as a product of zero-norm fields 
breaks the chiral symmetry explicitly (unless the mass of the PV fermion 
is taken to be zero), and we have been careful to notice whether or not 
it is restored in the large-PV-mass limit, at least in the sense that 
the physical mass of the fermion is proportional to the bare mass.  We 
find that for some ways of taking the limit the discrete chiral symmetry
is restored and for some ways it is not.  We do not know whether this consideration 
can provide a valid way of choosing one limiting procedure over another.  This 
question is important because, not only is chiral symmetry of 
interest in itself, but the way it is broken by the regulation procedure 
is analogous to the way gauge symmetry is broken by writing the interactions 
of gauge theories as products of zero-norm fields.

We have argued that our results suggest that if calculations are done 
in a truncated representation space, it may not be correct to take the 
limit of the PV masses going all the way to infinity.  It is easy to 
understand the reason why: if we are to have accurate calculations, most 
of the support of the wave functions in which we are interested must lie in the 
part of the space we retain.  We know from past studies that the projection 
of the low-lying states onto the higher Fock sectors often falls off very 
rapidly in the light-cone representation, but those results were for finite 
values of the regulators.  At infinite values of the regulators, the 
eigenvectors are not expected to exist at all, and we must expect that 
as the regulators are removed the projection of the wave functions onto 
any allowed sectors will become large. Thus it will be necessary to keep 
the PV masses finite when one truncates the representation space. If there 
are values of the PV masses sufficiently 
large to remove most of the bad effects of the negative-norm states on 
the eigenvectors in which we are interested, but small enough to make 
small the projection of these eigenvectors onto sectors we cannot manage to 
keep, then we can do a useful calculation; otherwise not.  We are currently 
performing studies to try to make these remarks quantitative.

\section*{Acknowledgments}
This work was supported by the Department of Energy
through contracts DE-AC03-76SF00515 (S.J.B.),
DE-FG02-98ER41087 (J.R.H.), and DE-FG03-95ER40908 (G.M.).
We also thank the Los Alamos National Laboratory for its hospitality
while this work was being completed.


\end{document}